%% file: main.tex
\documentclass[sigconf]{acmart}
\AtBeginDocument{%
  \providecommand\BibTeX{{%
    \normalfont B\kern-0.5em{\scshape i\kern-0.25em b}\kern-0.8em\TeX}}}

\acmSubmissionID{apfp1020}
\copyrightyear{2022}
\acmYear{2022}
\setcopyright{acmlicensed}\acmConference[KDD '22]{Proceedings of the 28th ACM SIGKDD Conference on Knowledge Discovery and Data Mining}{August 14--18, 2022}{Washington, DC, USA}
\acmBooktitle{Proceedings of the 28th ACM SIGKDD Conference on Knowledge Discovery and Data Mining (KDD '22), August 14--18, 2022, Washington, DC, USA}
\acmPrice{15.00}
\acmDOI{10.1145/3534678.3539080}
\acmISBN{978-1-4503-9385-0/22/08}

\DeclareMathOperator*{\argmax}{arg\,max}

\usepackage[norelsize, algo2e, ruled, lined, linesnumbered]{algorithm2e}
\usepackage{amsmath}
\usepackage{multirow}

\usepackage{subcaption}

\SetAlFnt{\small}
\SetKw{KwBy}{by}

\usepackage[belowskip=4pt,aboveskip=0pt]{caption}

\newtheorem{definition}{Definition}

\begin{document}

\title{TwHIN: Embedding the Twitter Heterogeneous Information Network for Personalized Recommendation}

\author{Ahmed El-Kishky$^\mathbf{*}$}
\affiliation{%
  \institution{Twitter Cortex}
  \city{Seattle}
  \state{WA}
  \country{USA}
}
\author{Thomas Markovich}
\affiliation{%
  \institution{Twitter Cortex}
  \city{Boston}
  \state{MA}
  \country{USA}
}

\author{Serim Park}
\affiliation{%
  \institution{Twitter Cortex}
  \city{San Francisco}
  \state{CA}
  \country{USA}
}

\author{Chetan Verma}
\affiliation{%
  \institution{Twitter Cortex}
  \city{San Francisco}
  \state{CA}
  \country{USA}
}

\author{Baekjin Kim}
\affiliation{%
  \institution{Twitter}
  \city{San Francisco}
  \state{CA}
  \country{USA}
}

\author{Ramy Eskander}
\affiliation{%
  \institution{Twitter Cortex}
  \city{New York}
  \state{NY}
  \country{USA}
}

\author{Yury Malkov}
\affiliation{%
  \institution{Twitter Cortex}
  \city{San Francisco}
  \state{CA}
  \country{USA}
}

\author{Frank Portman}
\affiliation{%
  \institution{Twitter Cortex}
  \city{Boston}
  \state{MA}
  \country{USA}
}

\author{Sof\'ia Samaniego}
\affiliation{%
  \institution{Twitter Cortex}
  \city{San Francisco}
  \state{CA}
  \country{USA}
}

\author{Ying Xiao$^\mathbf{\dagger}$}
\affiliation{%
  \institution{Twitter Cortex}
  \city{San Francisco}
  \state{CA}
  \country{USA}
}

\author{Aria Haghighi$^\mathbf{\dagger}$}
\affiliation{%
  \institution{Twitter Cortex}
  \city{Seattle}
  \state{WA}
  \country{USA}
}

\thanks{$^\mathbf{*}$ Corresponding author: aelkishky@twitter.com}
\thanks{$^\mathbf{\dagger}$ Equal contribution}

\renewcommand{\shortauthors}{Ahmed El-Kishky et al.}

\input{00abstract}



\maketitle

\input{01intro}

\input{02related}

\input{03preliminaries}

\input{04embedding_architecture}

\input{05recommender}

\input{06experiments}

\input{07design}

\input{08conclusion}

\balance

\bibliographystyle{ACM-Reference-Format}
\bibliography{main}

\appendix

\end{document}

%% file: 00abstract.tex
\begin{abstract}

Social networks, such as Twitter, form a heterogeneous information network (HIN) where nodes represent domain entities (e.g., user, content, advertiser, etc.) and edges represent one of many entity interactions  (e.g, a user re-sharing content or ``following'' another). Interactions from multiple relation types can encode valuable information about social network entities not fully captured by a single relation; for instance, a user's preference for accounts to follow may depend on both user-content engagement interactions and the other users they follow. In this work, we investigate knowledge-graph embeddings for entities in the Twitter HIN (TwHIN); we show that these pretrained representations yield significant offline and online improvement for a diverse range of downstream recommendation and classification tasks: personalized ads rankings, account follow-recommendation, offensive content detection, and search ranking. We discuss design choices and practical challenges of deploying industry-scale HIN embeddings, including compressing them to reduce end-to-end model latency and handling parameter drift across versions. 
\end{abstract}

%% file: 01intro.tex
\section{Introduction}
\begin{figure}[t]
\vspace{0.3cm}
    \centering
    \includegraphics[scale=0.5]{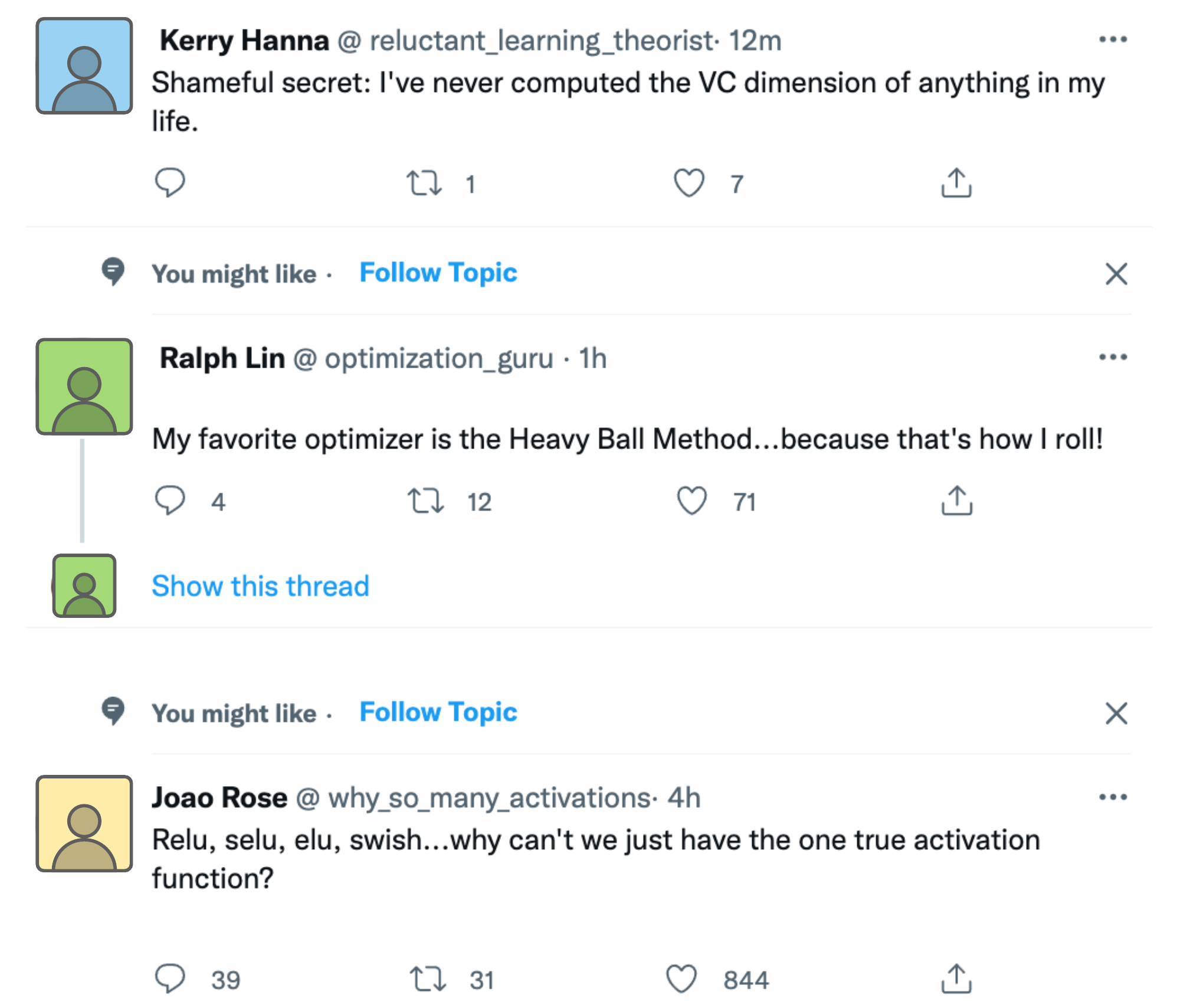}
    \caption{A mock-up of a Twitter feed. Notice the ``Reply'', ``Retweet'' and ``Favorite'' icons at the bottom of each Tweet.}
    \label{fig:recommendation}
\end{figure}

Twitter is an online social network where users post short messages called Tweets. When users visit Twitter, they can perform a variety of actions -- apart from ``Favoriting'', ``Replying'' and ``Retweeting'' tweets (Figure~\ref{fig:recommendation}); users can also ``Follow'' other users, search for tweets, click on ads, and open personalized notifications sent to mobile devices. With hundreds of millions of users, and billions of interactions per day, it is a challenging machine learning task to develop holistic (i.e., leverage all our data, independent of modality) and general representations that allow us to understand user preferences and behaviours across the entire platform. Such an understanding is critical in any number of personalized recommendation tasks at Twitter, and for providing compelling user experiences.

Large pretrained embedding tables\footnote{We make the distinction between \emph{pretrained} parameters, which are built independently of a downstream task, and \emph{trainable} embeddings, which are tuned as part of end-to-end task training. We focus on the former case here and do not consider cases of \textit{fine-tuning} where pretrained parameters are used to initialize task parameters, but tuned discriminatively per task.} have become an important tool in recommender systems; they have allowed ML practitioners to understand  user behavior. When pretrained embeddings are used as features, they have been shown to improve ranking models~\cite{cheng2016wide,covington2016deep,ying2018graph,you2019hierarchical}. In particular, pretrained user embeddings have been used in a variety of industry recommender systems such as for app and video recommendation for Google Play and Youtube~\cite{cheng2016wide,covington2016deep}, personalized search at AirBnB~\cite{grbovic2018real}, pin recommendation at Pinterest~\cite{ying2018graph, you2019hierarchical, pal2020pinnersage}, connecting users based on interest at Etsy~\cite{zhao2018learning}, stance detection at Twitter~\cite{pouguebiyong2022learning}, and news article recommendation at Yahoo! Japan~\cite{okura2017embedding}. 

The above methods have largely focused on utilizing user-item relations to learn embeddings for an associated recommendation task. For example, \textsc{PinSage}~\cite{ying2018graph} exclusively utilizes the ``click'' or ``repin'' actions that Pinterest users take on pins (i.e, content items) to create user embeddings for pin recommendation. While these approaches are successful for the item recommendation task for which they were designed, it may not generalize or be applicable to related recommendation tasks.
Returning to the example of Pinterest, there are other entity-relations beyond the user-pin interactions which may be relevant for recommendations. For instance, a user choose to "follow" a variety of other entity-types (another user, board, or topic). A representation which capture all these entity-type relations may be more useful in downstream tasks (such as user-topic or user-board recommendations) than representations learned from user-item interactions alone. These follow relations may also benefit item recommendation, since they are indications of the type of content a user is seeking. In general, leveraging a rich diversity of relations between entities can have many significant advantages to learning representations across many tasks simultaneously:

\begin{description}
\item[Data Supplementation:] For some tasks, there may simply be fewer data points for training models (e.g., there are generally fewer ad than organic content engagements, or a new product feature may have low density of interactions). In these cases, supplementing low-density relations with information from ``denser'' relations in the network may improve predictive ability of embeddings derived from the network. 
\item[Task Reusability:] Often-times, we do not even know all the downstream tasks ahead of time; building `universal' representations reduces the labour-intensive process of identifying, training, and managing multiple embeddings.
\end{description}
 
 To address these weaknesses, we model multi-type multi-relational networks at Twitter as a heterogeneous information network (HIN)~\cite{sun2013mining,shi2016survey}, and apply scalable techniques from knowledge graph embeddings (KGE) to embed our heterogeneous networks~\cite{bordes2013translating,trouillon2016complex,lerer2019pytorch}.

Much of the literature in this area either focuses on small-scale embeddings without deploying models to production or industry-scale recommender systems that are trained on simple networks with only a few distinct types of entities and relationships, thereby limiting the embedding's utility to a small number of applications. In this work, we present an end-to-end outline of embedding the \textbf{Tw}itter \textbf{H}eterogeneous \textbf{I}nformation \textbf{N}etwork (TwHIN). Our end-to-end system is deployed in production at Twitter, across a variety of product areas; training is scalable, operating on more than $10^{9}$ nodes and $10^{11}$ edges, and can incorporate many disparate network sources for richer embeddings. TwHIN embeddings capture signals such as social signals (follow-graph), content engagement signals (Tweet,  image, and video engagement), advertisement engagements, and others to learn embeddings for users, tweets, and advertisements and other types. 
We evaluate TwHIN embeddings in online (A/B) tests and offline experiments, demonstrating improvement in multiple tasks.

Compared to previous papers on learning industry-scale embeddings for Web recommender systems, our contributions are:

\begin{itemize}
    \item We demonstrate simple KGE techniques offer a scalable, flexible scheme for embedding large heterogeneous graphs. This is in contrast to previous industry efforts~\cite{ying2018graph} which require complex alternating CPU-GPU rounds, multiple MapReduce jobs, and custom  OpenMP extensions; the KGE embeddings we use need only a single multi-GPU machine.
    \item We demonstrate how heterogeneous embedding approaches combine disparate network data to effectively leverage rich, abundant unlabeled data sources to learn better representation while simultaneously addressing data sparsity. We show this approach yields gains in multiple downstream recommendation and classification tasks
    \item We detail practical insights, design considerations and learnings in developing and productionizing a single heterogeneous network embedding for use in a variety of disparate recommender systems within Twitter.
\end{itemize}

%% file: 02related.tex
\section{Related works}
\label{sec:related}
Network embedding (a.k.a., graph embedding) techniques have been proposed as a means to represent the nodes of large networks into low-dimensional dense representations~\cite{hoff2002latent,yan2005graph}. The information contained within these node embeddings has proven useful in a variety of data mining tasks including classification, clustering, link prediction, and recommendation~\cite{goyal2018graph,cai2018comprehensive,wei2017cross}. A family of approaches, starting from \textsc{DeepWalk}, provide a scalable approach to graph embedding by performing a random walk to create a node sequence, and then applying SkipGram to learn node embeddings~\cite{perozzi2014deepwalk,tu2016max}. Node2vec extended DeepWalk by applying a biased random walk procedure with a controllable parameter between breadth-first and depth-first sampling strategies~\cite{grover2016node2vec}. Later methods such as LINE, SDNE, and GraRep incorporate second and higher-order proximity in the node embedding objective~\cite{tang2015line,wang2016structural,cao2015grarep}. More complex approaches have applied stacked denoising autoencoders to learn node embeddings~\cite{cao2016deep}. Despite the plethora of popular network embedding techniques, these methods largely cater to homogeneous networks -- those with a single type of edge relation -- with no clear adaptation to type-rich heterogeneous networks.

Generalizing beyond homogeneous networks, heterogeneous information networks have been proposed as a formalism to model rich multi-typed, multi-relational network data~\cite{shi2016survey,sun2013mining,xin2018active,wang2015incorporating}. In this setting, one common use-case has been to perform similarity computation between nodes based on structural similarities; several path-based methods have been proposed for this similarity search~\cite{sun2011pathsim,shi2014hetesim}. Recognizing the utility of HINs in recommendations, there have been many approaches to combining these two ideas. One method addresses the cold-start problem by incorporating heterogeneous network data in social tagging~\cite{feng2012incorporating}. Another work uses heterogeneous relations in a collaborative filtering social recommendation system~\cite{luo2014hete}. Other approaches have exploited multi-hop meta-paths over HINs to develop collaborative filtering models for personalized recommendation~\cite{shi2015semantic}. 
Additional methods have used these multi-hop meta-paths as latent features within recommender systems~\cite{yu2013recommendation}. Meta-path similarities were later applied as regularization within matrix factorization recommender systems~\cite{yu2013collaborative}. Follow-up works have leveraged the rich plethora of entity relationships to perform personalized recommendation~\cite{yu2014personalized}. Our approach differs from these methods in that we leverage the plethora of heterogeneous relationships to learn better entity representations (embeddings); these embeddings can then be directly incorporated as dense features in state-of-the-art deep-learning-based recommender models.

Instead of directly operating on the HIN for recommender systems, several papers have investigated learning representations from heterogeneous networks. Several approaches have been developed to learn content-enriched node representations~\cite{zhang2016homophily,yang2015network}. These methods address data sparsity by leveraging all content associated with nodes. Other methods have been developed to directly embed nodes in a HIN~\cite{chang2015heterogeneous,tang2015pte,xu2017embedding,chen2017task,dong2017metapath2vec}. This work is the closest to our task, however many of these techniques do not easily scale to industry-scale networks such as TwHIN; additionally, several embedding techniques are custom-derived for specific network schema. As an alternative, we apply knowledge graph embedding (KGE) techniques to embed our networks~\cite{wang2017knowledge,bordes2013translating,trouillon2016complex}. As KGEs  can incorporate mutli-typed nodes and edges, they translate naturally to embedding HINs. Additionally, several frameworks have been developed to scale KGEs to billions of nodes and trillions of edges~\cite{lerer2019pytorch,DGL-KE,zhu2019graphvite}.

%% file: 03preliminaries.tex
\section{Preliminaries}
\label{sec:preliminaries}
Twitter contains a plethora of multi-typed, multi-relational network data. For example, users can engage with other users (i.e., the `following' relation), which forms the backbone of the social follow-graph. Additionally, users can engage with a variety of non-user entities (e.g., Tweets and advertisements) within the Twitter environment using one of several engagement actions (Favorite, Reply, Retweet, Click). We model these networks as \textit{information networks} ~\cite{sun2011pathsim,yu2014personalized}:

\begin{definition}[Information Network]
\label{def:hin}
An information network is defined as a directed graph $G=(V,E, \phi, \psi)$ where $V$ is the set of nodes, $E$ is the set of edges, $\phi$ is an entity-type mapping function ($\phi: V \rightarrow \mathcal{T}$) and $\psi$ is an edge-type mapping function ($\psi: E \rightarrow \mathcal{R}$). Each entity $v \in V$ belongs to an entity type $\phi(v) \in \mathcal{T}$, and each edge $e \in E$ belongs to a relation type $\psi(e) \in \mathcal{R}$
\end{definition}

\begin{figure}
    \centering
    \includegraphics[scale=0.7]{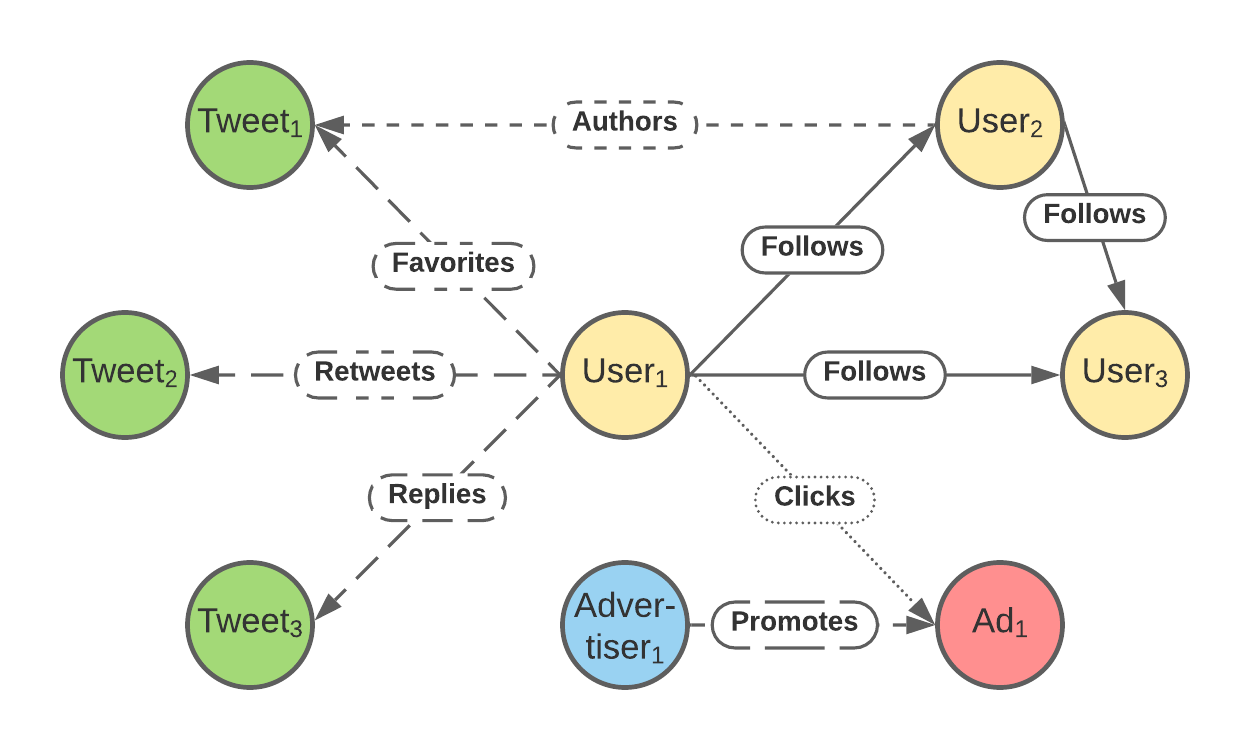}
    \caption{An example heterogeneous information network (HIN) where $|V| = 8$ and $|E|=9$. There are four entity types ($\mathcal{T}$): `User', `Tweet', `Advertiser',  and `Ad'. There are seven types of relationship ($\mathcal{R}$): `Follows', `Authors', `Favorites', `Replies', `Retweets', `Promotes', and `Clicks'. See Section~\ref{sec:preliminaries} for more details.}
    \label{fig:twhin-schema}
\end{figure}

An information network is a \textit{heterogeneous information network} when $|\mathcal{T}| > 1$ or $|\mathcal{R}| > 1$. For consistency with recommender system terminology, we refer to entities being recommended as \textit{items}. In Figure~\ref{fig:twhin-schema}, we give a small example HIN.

Given an input HIN, $G$ and an input dimension $d$, our goal is to learn embeddings for each entity in $G$. We define these HIN embeddings as follows:

\begin{definition}[HIN Embeddings]
\label{def:hin-embedding}
Given a network $G = (V, E, \phi, \psi)$, the heterogeneous information network embedding uses self-supervised structure prediction tasks in $G$ to map entities in $V$ and relations in $\mathcal{R}$ onto a low-dimensional space $\mathbb{R}^d$, where $d \ll |V|$.
\end{definition}

In particular, our goal is to learn HIN embeddings that provide utility as features in downstream recommendation tasks. 

%% file: 04embedding_architecture.tex
\section{TwHIN embeddings}
\label{sec:embeddings}
In this section, we describe our approach to extracting information from the rich multi-typed, multi-relation Twitter network through large-scale knowledge-graph embedding. We then describe our use of clustering to inductively infer multiple embeddings for each user, and out-of-vocabulary entities such as new tweets without retraining on the new graph. Finally, we describe the overall end-to-end scheme, from the raw data sources to downstream task training.

\subsection{HIN embedding approach}
\label{sec:objective}
We apply knowledge graph embedding to embed the Twitter HIN (TwHIN)~\cite{bordes2013translating,trouillon2016complex,lin2015learning,wang2014knowledge}. We represent each entity, as well as each edge type in a HIN as an embedding vector (i.e., vector of learnable parameters). We will denote this vector as $\theta$. A triplet of a source entity $s$, edge type $r$, and target entity $t$ is scored with a scoring function of the form $f(\theta_s, \theta_r, \theta_t)$. Our training objective seeks to learn $\theta$ parameters that maximize a log-likelihood constructed from the scoring function for $(s, r, t) \in E$ and minimize for $(s, r, t) \notin E$.

For model simplicity, we apply translating embeddings (\textsc{TransE}) to embed entities and edges in a HIN~\cite{bordes2013translating}. For an edge $e=(s, r, t)$, this operation is defined by:
\begin{equation}
    \label{eq:scoring}
    f(e) = f(s, r, t) = (\theta_s + \theta_r)^\intercal \theta_t
\end{equation}

As seen in Equation~\ref{eq:scoring}, \textsc{TransE} operates by translating the source entity's embedding with the relation embedding; the translated source and targets embeddings are then scored with a simple scoring function such as a dot product.

We formulate the learning task as an edge (or link) prediction task. We consume the input HIN $G$ as a set of triplets of the form $(s, r, t)$ which represent positive, observed edges. The training objective of the translating embedding model is to find entity representations that are useful for predicting which other entities directly are linked by a specific relation. While a softmax is a natural formulation to predict a linked entity, it is impractical because of the prohibitive cost of computing the normalization over a large vocabulary of entities. As such, following previous methods~\cite{mikolov2013distributed,goldberg2014word2vec},  negative sampling, a simplification of noise-contrastive estimation, is used to learn the  parameters $\theta$. We maximize the following negative sampling objective: 
\begin{equation}
    \label{eq:objective}
    \argmax_{\theta}\sum_{e \in G} \left[  \log \sigma (f(e)) + \sum_{e' \in N(e)} \log \sigma (-f(e')) \right]
\end{equation}
where: $N(s, r, t) = \{(s, r, t'): t'\in V\} \cup \{(s', r, t): s\in V\}$.
Equation~\ref{eq:objective} represents the log-likelihood of predicting a binary ``real" or ``fake" label for the set of edges in the network (real) along with a set of the "fake" negatively sampled edges. To maximize the objective, we learn $\theta$ parameters to differentiate positive edges from negative, unobserved edges. Negative edges are sampled by corrupting positive edges via replacing either the source or target entity in a triple with a negatively sampled entity of the same entity type. As input HINs are very sparse, randomly corrupting an edge in the graph is very likely to be a 'negative' edge absent from the graph. 
Following previous approaches, negative sampling is performed both uniformly and proportional to entity prevalence in the training HIN~\cite{bordes2013translating,lerer2019pytorch}. Optimization is performed via Adagrad \cite{duchi2011adaptive}.

\subsection{Computational considerations}\label{sec:computational}
 As an input HIN for Twitter can include more than $10^9$ nodes such as Users, Tweets, and other entities, learning an embedding vector for each entity presents both system and GPU memory challenges. We apply the framework PyTorch-Biggraph~\cite{lerer2019pytorch} to address the large memory footprint. This framework randomly partitions using a partition function $\pi$ each node $v$ into one of $P$ partitions ($\pi(v) \in \{0, \ldots P{-}1\}$), where $P$ was selected based on the memory available on the server and the GPUs for training. As each node is allocated to a partition, edges $e=(s,r,t)$ are allocated to buckets based on the the source and target nodes ($s$ and $t$). As such the partitions form $P^2$ buckets and an edge falls into bucket $B_{\pi(s),\pi(t)}$. Buckets of edges and their associated source and target entities' embedding tables are loaded into memory and embeddings are trained. As such, a maximum of approximately $2V/P$ entities' embeddings are loaded into memory at any point. 
 \begin{algorithm2e}
\caption{HIN Embedding}
\label{alg:embedding}
\Indm
      \KwIn{$G=(V, E, \phi, \psi), epochs, P $}
      \KwOut{$\theta$ (learned embeddings)}
\Indp

      \BlankLine
    let $\theta$: initialized embedding vectors entities in $V$ relations in $\mathcal{R}$  \\
    \For{each $\{1 \ldots epochs\}$}{
        \For{each bucket $(i, j): 0 {\le} i {<} P;~0 {\le} j {<} P$ )}{
        load bucket $B_{i,j}$ edges onto memory \\
        load $\{\theta_v: \pi(v)=i \lor \pi(v)=j\}$ onto GPU\\
        train embeddings on edges using Equation~\ref{eq:scoring} \\
        }
	}	
\end{algorithm2e}

Algorithm~\ref{alg:embedding} is then applied to scalably learn $\theta$. The algorithm simply selects a random bucket and loads the associated partitions' embedding tables onto GPU memory. Negative examples for each bucket edge are sampled as described above, but limited only to entities present in the current buckets. Gradients from the edge prediction among negative samples task is backpropagated to learn appropriate embedding vectors. 

\subsection{HIN at Twitter: TwHIN}
When applying Algorithm \ref{alg:embedding}, we must take care to make a crucial distinction between relation types. Within TwHIN we identify relations that are high-coverage in the number of users that participate in the relation, and contrast them to low-coverage relations that are overall sparse. For example, most users follow at least one other user and engage with at least a small number of Tweets. However, many users may not engage with advertisements at all. Recognizing this distinction, we ensure that high-coverage relations are co-embedded with low-coverage relations, but not with other high-coverage relations. This ensures that high-coverage relations benefit entities in low-coverage relations by addressing sparsity, while preserving entities in high-coverage relations from cross-entity type interference.

We form two heterogeneous networks centered around the two high-coverage relations: (1) follow-graph and (2) User-Tweet engagement graph. We refer to these two networks as TwHIN-Follow and TwHIN-Engagement. For each network we augment the high-coverage relation with low-coverage relations with entities such as promoted Tweets (Advertisements), advertisers, and promoted mobile apps.  We performed TwHIN embedding on a single computer which allowed for fast prototyping, experimentation, and productionization. We used the Pytorch BigGraph\footnote{\url{https://github.com/facebookresearch/PyTorch-BigGraph}} framework to perform embedding at scale~\cite{lerer2019pytorch}.
\begin{figure*}[t]
    \centering
    \includegraphics[scale=0.65]{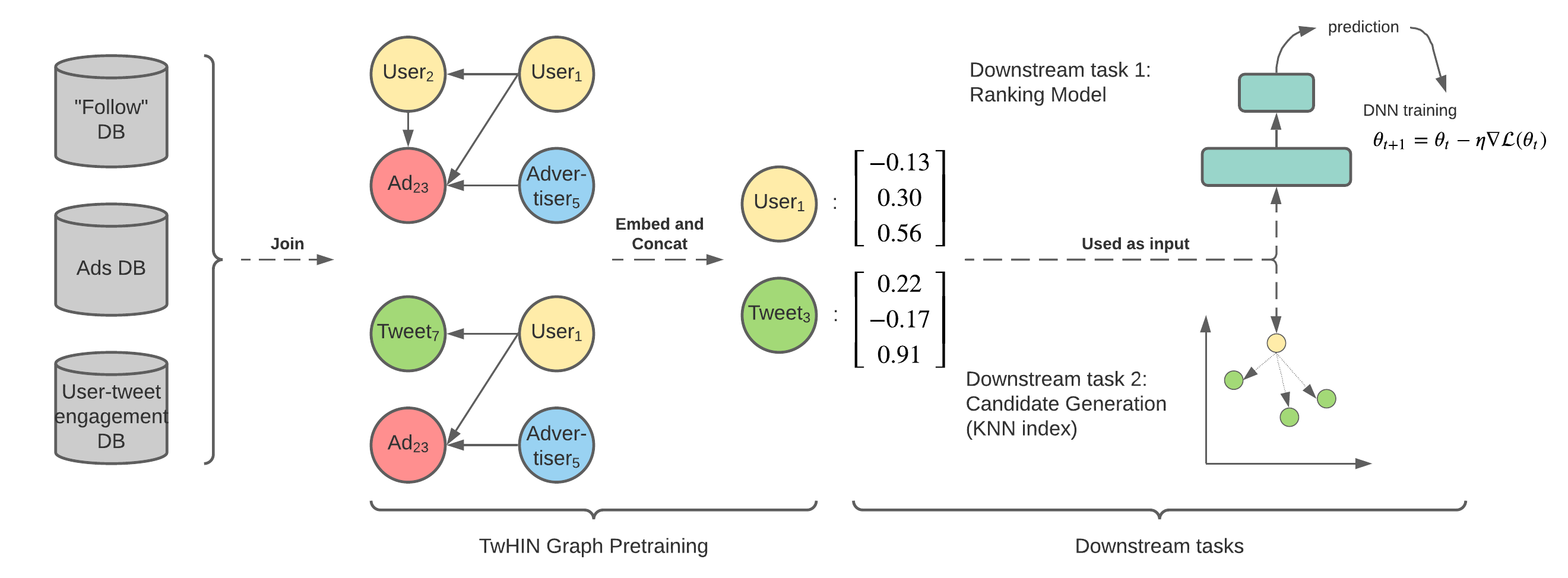}
    \caption{The end-to-end framework aggregates disparate network data to construct TwHIN, joint-embedding is performed and embeddings are consumed in downstream tasks and ML models.}
    \label{fig:architecture}
\end{figure*}

\subsection{Inductive multi-modal embeddings}
\label{sec:inductive}
While classical knowledge graph embedding techniques provide a scalable way to embed large HINs, they suffer from two major shortcomings:  (1) entites are represented by only a single embedding, which can fail to capture complex behaviour where a user might have disparate multi-modal interests  (2) entity embeddings are \textit{transductive} and only defined for entities present at training time and retraining is required on the full graph to support any novel out-of-vocabulary nodes. This latter shortcoming is significant for application to TwHIN since the set of Tweets and Users change rapidly over time.

We address both of these short-comings by introducing a fast post-processing step that can represent TwHIN entities as mixture over multiple embeddings. This technique is flexible enough to \textit{inductively} embed new, out-of-vocabulary, nodes such as Tweets. To create these embeddings for a node type we (1) cluster existing unimodal embeddings (2) compute multiple embeddings for a node by aggregating the most engaged-with clusters for a node. 

To illustrate this approach, let us focus on transforming User embeddings into multi-modal mixtures of embeddings. We first consider set $T$ of non-user entities that have some engagement with users; we think of $T$ as the set of ``targets`` and will seek embeddings for users that summarize a user's engagement with elements of $T$. We first take the set of targets and perform k-means clustering~\cite{sculley2010web}  to create a set of  clusters $C$ over $T$. We then associate each user $u_i$ with a probability distribution over elements of $C$ proportional to the amount of engagement the user has with each cluster:

\begin{equation}
    P(c|u_i) = \frac{\mathrm{count}(u_i,c)}{\sum\limits_{c'\in \mathcal{M}(u_i)} \mathrm{count}(u_i, c')}
    \label{eq:distribution}
\end{equation}

Where, $\mathrm{count}(u_i, c)$ is the number of times $u_i$ engages with a target in cluster $c$ and, for computational efficiency, we take $\mathcal{M}(u_i)$ to be $u_i$'s top $m$ most engaged clusters. We normalize these cluster engagements to obtain a proper cluster-engagement distribution. In addition to each user's TwHIN entity embedding, we can now represent each user with a mixture distribution over their top $m$ most engaged cluster with each cluster represented by a dense vector of its centroid or medoid. This multi-modal representation addresses both of the short-comings since different clusters of target entities may better capture complex behavior and they can also be generated for entities that were unseen during training.

The multi-modal embeddings here have similar motivations as those in \textsc{PinnerSage}~\cite{pal2020pinnersage}, but have some key differences. In that work a user's target engagements are clustered and individual items capturing disparate interests are used to represent the user for recommendation retrieval. By contrast, in the approach here, we are clustering the entire universe of items (Tweets and other non-user entities) and representing users according to which of these global clusters their engagements coincide. So rather than run clustering for each user on just their engagement, we instead run a single large clustering job over non-user entities.

%% file: 05recommender.tex
\section{TwHIN for recommendation}
\label{sec:recsys}
In Figure \ref{fig:architecture}, we show our end-to-end framework including collecting disparate data sources to organize TwHIN, learning entity embeddings through a self-supervised KGE objective, and finally using the TwHIN embeddings in downstream tasks. In this section 
we discuss using TwHIN embeddings for two families of tasks: (1) candidate generation and (2) as features in deep-learning models for recommendation and prediction. 

\subsection{TwHIN candidate generation}
\label{sec:cand_gen}
Candidate generation is the first step in most recommender systems; the aim is to retrieve a user-specific high-recall set of relevant candidates using a light-weight strategy. Within Twitter, we use an approximate nearest neighbor (ANN) system that indexes items to be suggested such as Users to Follow or Tweets to engage with. Two  internal systems use HNSW~\cite{malkov2018efficient} or FAISS~\cite{JDH17} with Product Quantization~\cite{jegou2010product} to index items and retrieve candidates.

We then use an entity's TwHIN embedding to query candidate entities of any type (assuming a distinct index per entity-type). However, when indexing a large number of items such as users or Tweets, many of the retrieved items may be very similar. These  are not desirable as users get diminishing value from being presented near duplicate items. To address this, we use multi-modal embeddings (see Section~\ref{sec:inductive}) to generate diverse candidates. Given an entity's multimodal representation as a \textit{mixture over multiple embeddings} with non-negative mixture coefficients normalized to one, we can query candidates from each vector in its mixture representation and select a number of candidates proportional to the query vector's mixture weights. This adds diversity as candidates are explicitly queried from different areas of the embedding space. 

\subsection{TwHIN for ranking and prediction}
\label{sec:ranking}
Many supervised machine learning models at Twitter employ pretrained TwHIN embeddings as features. These models are used in a variety of tasks from recommendations ranking, to content classification, and other predictive tasks. Many standard deep neural network (DNN) models have been applied to predictive modeling~\cite{cheng2016wide,wang2017deep,wang2021dcn}. Recommendation ranking models typically take as input a set of user and contextual information, and output is a ranked list of items from the candidate generation step based on objectives such as engagement or purchase probability. Predictive classification models take similar features and predict a classification label such as topic or other content classification.

At Twitter, predictive models take in users and/or items as described by many continuous and categorical features. Categorical features are then associated with an emebdding vector via a look-up table; these embedding vectors are task-specific and learned while training each DNN model. Continuous features are then appended to these embeddings and the concatenated feature-set is fed into a DNN (e.g., MLP) and trained on a task-specific objective. To incorporate TwHIN embeddings, we employ a look-up table to map an entity id to its associated pretrained TwHIN embedding. However, unlike with other categorical features, these pretrained embeddings are frozen and not trained with the unfrozen embeddings.

%% file: 06experiments.tex
\vspace{-0.15cm}
\section{Experiments \& results}
\label{sec:exps}
We experimentally demonstrate the generality and utility of TwHIN embeddings via online and offline experimentation on several Twitter internal ML models and tasks.

\subsection{Candidate generation}
\label{sec:exp:cand_gen}
Our first family of tasks are \textit{candidate generation} tasks; these are tasks that select a high-recall pool of items that are then ranked by more complex downstream ranking models. We demonstrate offline and online gains in using TwHIN embeddings on a ``Who to Follow'' task.

\ \\
\noindent
\textbf{Who to Follow Suggestions:}
We describe results from leveraging TwHIN embeddings for the \textit{Who to Follow}~\cite{gupta2013wtf,el2022knn} user recommendation task which suggests Twitter accounts for a user to follow. We utilize TwHIN user embeddings as a query to retrieve highly followed users via approximate nearest neighbor search. We compare using a single user embedding for querying to using multi-modal ``mixture of embeddings'' as described in Section~\ref{sec:inductive} and Section~\ref{sec:cand_gen}.

\begin{table}[h]
 \centering
    \begin{tabular}{l c c c}
    \toprule
    \textbf{Approach} & \textbf{R@10} & \textbf{R@20} & \textbf{R@50}\\
    \midrule
    Unimodal & 0.58\% & 1.02\%  & 2.06\%  \\
    Mixture & 3.70\% & 5.53\%  & 8.79\%  \\
    \bottomrule
    \end{tabular}
    \caption{Comparing candidate generation using a single TwHIN embedding vs a mixture of embeddings with multi-querying.}    
    \label{tab:wtf}
\end{table}

Table~\ref{tab:wtf} compares the performance of multi-modal mixtures of TwHIN user embeddings vs unimodal user embeddings. Across all thresholds (recall at $10$, $20$, and $50$), multi-modal mixtures with multi-querying significantly outperform single unimodal representations. This confirms the hypothesis that mixtures of embeddings better model users and their multiple interests. Explicitly querying from different parts of the embedding space consistently yielded over $300\%$ improvement in recall over unimodal representation querying.

\subsection{Recommendation and prediction}
\label{sec:exp_ranking}
We incorporate TwHIN embeddings in several supervised recommendation and prediction tasks: (1) \textit{Predictive Advertisement Ranking} (2) \textit{Search Ranking}, and (3) \textit{Offensive Content Detection}

\ \\
\noindent
\textbf{Predictive Advertisement Ranking:}
Following procedures in Section~\ref{sec:recsys}, we add TwHIN embeddings for different entities to several ads recommendation models~\cite{o2021analysis}. We are unable to disclose specific details of each model, or the timeline in which they were deployed. As such, we simply refer to these models as $Ads_1$, $Ads_2$ and $Ads_3$; each targets a different ads target objective, but does share some, but not all, hand-crafted features. In particular, each model has many features about users and their interactions with ads.  

We evaluate our model quality using Relative Cross Entropy (RCE). This metric measures how much better an ads engagement predictor is compared to a naive prediction using the prior of each label. This is computed as follows:

\begin{equation}
    \label{eq:rce}
    \textrm{RCE} = 100 \times \frac{\textrm{Reference Cross Entropy} - \textrm{Cross Entropy}}{\textrm{Reference Cross Entropy}}
\end{equation}
where the reference cross entropy is that of the prior, and the cross entropy term is that of the treatment.

During online A/B experiments, where the approaches were tested on live traffic, we computed the pooled RCE (using both control and treatment traffic) and noticed a significant improvement when adding TwHIN embeddings over the baseline model. Adding TwHIN embeddings yielded an average $\mathbf{2.38}$ RCE gain over the baseline resulting in a $10.3\%$ cost-per-conversion reduction in the new production model. These results allowed us to deploy TwHIN embeddings for several additional advertisement models which all demonstrated online improvement. To further validate our hypothesis as to the importance of heterogeneity in TwHIN, we perform offline entity ablation studies for a set of advertisement models. Note that in some of our experiments, we observe that corresponding online RCE increase is a lot more significant than what is measured offline due to differences between offline and online environment. Notwithstanding, the offline results are directionally monotonic with those measured online. In particular, any offline results should be used to compare utilizing different entity embeddings within the same model.

\begin{table}[h]
\small
    \centering
    \begin{tabular}{l c c c c c}
        \toprule
       \textbf{Model} & \textbf{Baseline} & $\mathbf{U}$ & $\mathbf{U{+}A}$ &  $\mathbf{U{+}T}$ & $\mathbf{U{+}A{+}T}$    \\\midrule
        
       {$\mathbf{Ads_1}$}  & 21.23 & 21.32 & 21.43 & 21.46 & \textbf{21.48} 
            \\\midrule
    {$\mathbf{Ads_2}$} & 13.53  & 13.61 & 13.54 & \textbf{13.63} & 13.59  
    \\\midrule
       {$\mathbf{Ads_3}$}  & 17.11 & 17.26 & 17.27 & \textbf{17.27} & 17.26
        \\\bottomrule
        
    \end{tabular}
    \caption{Offline RCE for ads models with feature ablation. We investigate performance when using TwHIN embeddings for User (U), Advertiser (A), and Target entity (T) such as app to install, video to watch, or advertisement to click.}
    \label{tab:ads_ablation}
\end{table}

As a general trend, we notice the most improvement arises from user embeddings. However, we do see further improvements when adding entity embeddings for Advertiser and the exact target (e.g., promoted app for installation, promoted video, or promoted Tweet). This supports our intuition that leveraging multi-type embeddings can improve downstream predictive models. 

We also ran experiments (not shown) where we directly embedded low-coverage relationships without augmenting with denser relationships. This yielded much lower improvements, and sometimes even model degradation. This further supports our claim that denser relations can supplement less dense relations.

\ \\
\noindent
\textbf{Search Ranking:}
We investigate using TwHIN embeddings to improving personalized search ranking. Personalized search ranking consists of a search session for a user where they provide an input query and the system ranks a set of Tweets based on engagement. Our baseline ranking system takes as input a large set of hand-crafted features that represent the underlying user, query and candidate Tweets. In addition, the input features also include the outputs of an mBERT~\cite{devlin2018bert} variant fine-tuned on in-domain query-Tweet engagements to encode the textual content of queries and Tweets. The hand-crafted and contextual features are fed into an MLP, where the training objective is to predict whether a Tweet triggers searcher engagement or not.

We augment this baseline model with three TwHIN embeddings: User embeddings from both follow-base ($\mathbf{U_f}$) and engagement-base ($\mathbf{U_e}$), as well as Author embeddings ($\mathbf{A}$).

\begin{table}[h]
\setlength{\tabcolsep}{3.5pt}
\small
    \centering
    \begin{tabular}{ l c c c c c c}
        \toprule
        \textbf{Metric} & \textbf{Baseline} & $+\mathbf{U_f}$ & $+\mathbf{U_e}$ &  $+\mathbf{A}$ & $+\mathbf{U_f+U_e}$ & $\mathbf{+U_f+U_e+A}$ \\\midrule
        MAP & 55.7 & 56.2 & 56.6 & 55.8 & 56.6 & \textbf{57.0} \\
        ROC & 57.9 & 58.6 & 59.0 & 57.9 & 59.0 & \textbf{59.6} \\
        \bottomrule
    \end{tabular}
    \caption{Search engagement-based ranking with TwHIN embeddings: TwHIN user embeddings, both follow-base ($\mathbf{U_f}$) and engagement-base ($\mathbf{U_e}$), and TwHIN author embeddings ($\mathbf{A}$)}
    \label{tab:search_ablation}
\end{table}

We train our models on search-engagement data. We fine-tune the hyper-parameters on search sessions from a held-out validation day and report ranking performance in Table~\ref{tab:search_ablation} for both MAP and averaged ROC using search sessions from a held-out test day.

As seen in Table~\ref{tab:search_ablation}, combining the three types of TwHIN embeddings as additional inputs to the baseline system yields the best ranking performance with relative error reductions of 2.8\% in MAP and 4.0\% in averaged ROC. Note also that TwHIN user embeddings ($\mathbf{U_f}$ and $\mathbf{U_e}$), either independently or in conjunction, yield performance gains, unlike TwHIN author embeddings ($\mathbf{A}$), which only help when combined with user embeddings. 

\ \\
\noindent
\textbf{Detecting Offensive Content:} We evaluate TwHIN embeddings for the task of predicting whether a Tweet is offensive or promotes abusive behavior.\footnote{See \texttt{help.twitter.com/en/rules-and-policies/abusive-behavior} for a fuller discussion of what is considered abusive behavior in Tweets.}  While the definition of this problem is purely concerned with the Tweet content, we hypothesize that a key element of interpreting intent of a Tweet is understanding the social context and community of the Tweet author. These experiments are purely academic, and TwHIN is not currently being applied to detecting offensive content at Twitter.

For our experimental purposes, we construct a baseline approach that fine-tunes a large-scale language model for offensive content detection using linear probing and binary categorical loss; we compare the performance of RoBERTa~\cite{roberta2019} and BERTweet~ \cite{nguyen2020bertweet} language model, the latter of which has been pretrained on Twitter-domain data. We evaluate on two collections of tweets where some tweets have been labeled ``offensive" or violating guidelines. The baselines leverage pretrained language models to embed the textual content. We supplement the stronger baseline by concatenating TwHIN author embedding to the language model content embedding; linear probing is used for fine-tuning. 

\begin{table}
    \centering
    \begin{tabular}{c c c c}
        \hline
         & \textbf{RoBERTa} & \textbf{BERTweet} & \textbf{+TwHIN-Author}   \\
        \hline 
        $\mathbf{Collection}_1$& 0.4123 & 0.4692 & \textbf{0.5161} \\
        $\mathbf{Collection}_2$ & 0.688 & \textbf{0.7274} & 0.7174 \\
        \hline
    \end{tabular}
    \caption{PR-AUC results for detecting offensive content. We compare the performance of content-based models to leveraging a TwHIN author embedding in addition to content.}
    \label{experiment_toxicity}
\end{table}

Results in Table~\ref{experiment_toxicity} confirm our hypothesis showing that adding TwHIN embedding increases PR-AUC by an average relative gain of 9.09\% on $\mathrm{Collection}_1$ with neutral results on $\mathrm{Collection}_2$, likely stemming from $\mathrm{Collection}_2$ containing a very high proportion of offensive tweets. This experimental result confirms unrelated relationships (e.g., Follows and Tweet engagements) can be used to pretrain user embeddings that can improve unrelated predictive tasks such as offensive content or abuse detection validating our claim on the generality of our TwHIN embeddings. While TwHIN was constructed without any data from this task, the learned embeddings were able to significantly improve performance on this task.

%% file: 07design.tex
\begin{figure*}[t]
     \centering
     \begin{subfigure}[b]{0.3\textwidth}
         \centering
         \includegraphics[width=\textwidth]{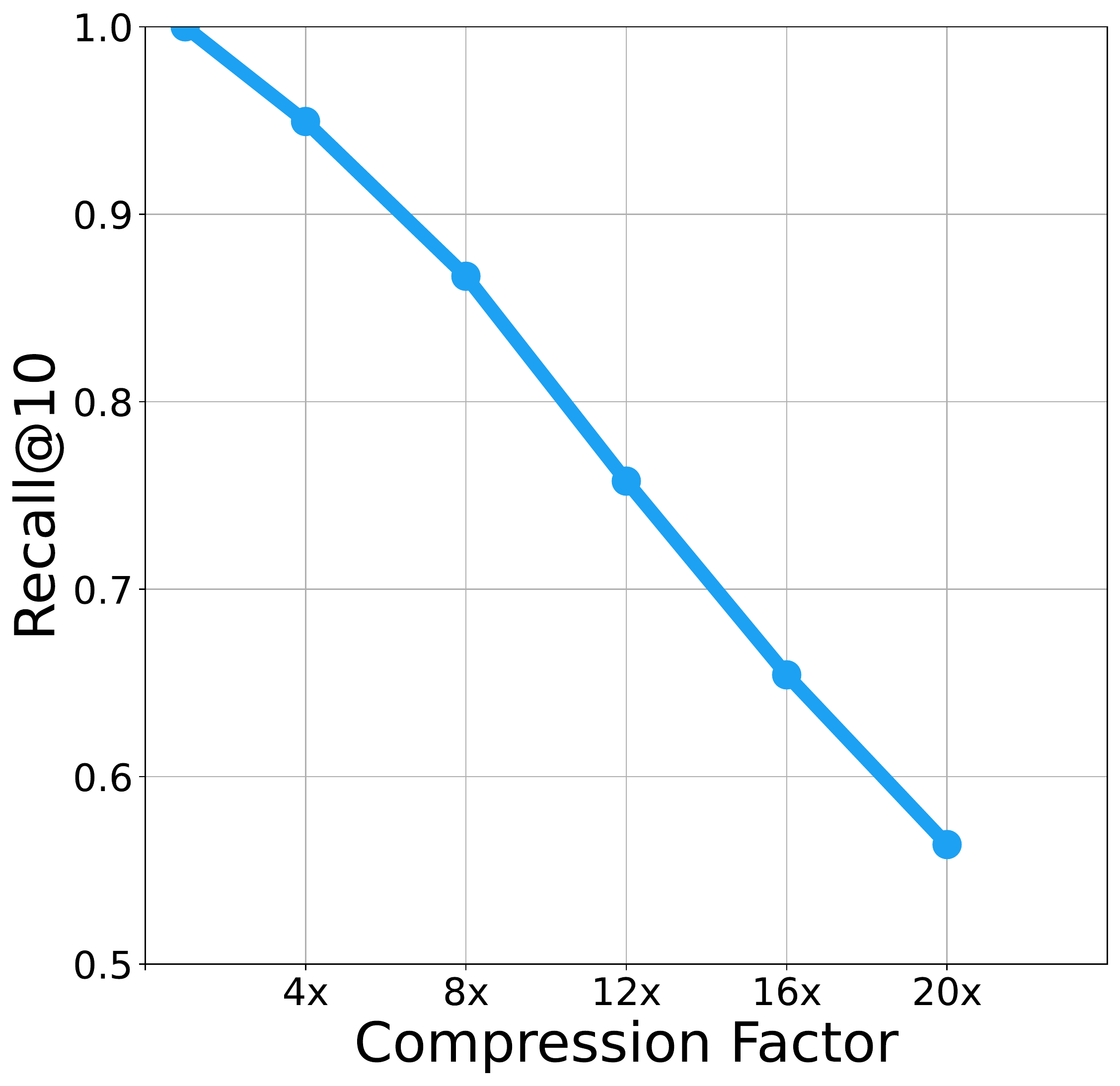}
         \caption{Recall@10 for $k$-nearest neighbors with varying levels of compression for TwHIN.}
         \label{fig:knn_compression}
     \end{subfigure}
     \hfill
     \begin{subfigure}[b]{0.315\textwidth}
         \centering
         \includegraphics[width=\textwidth]{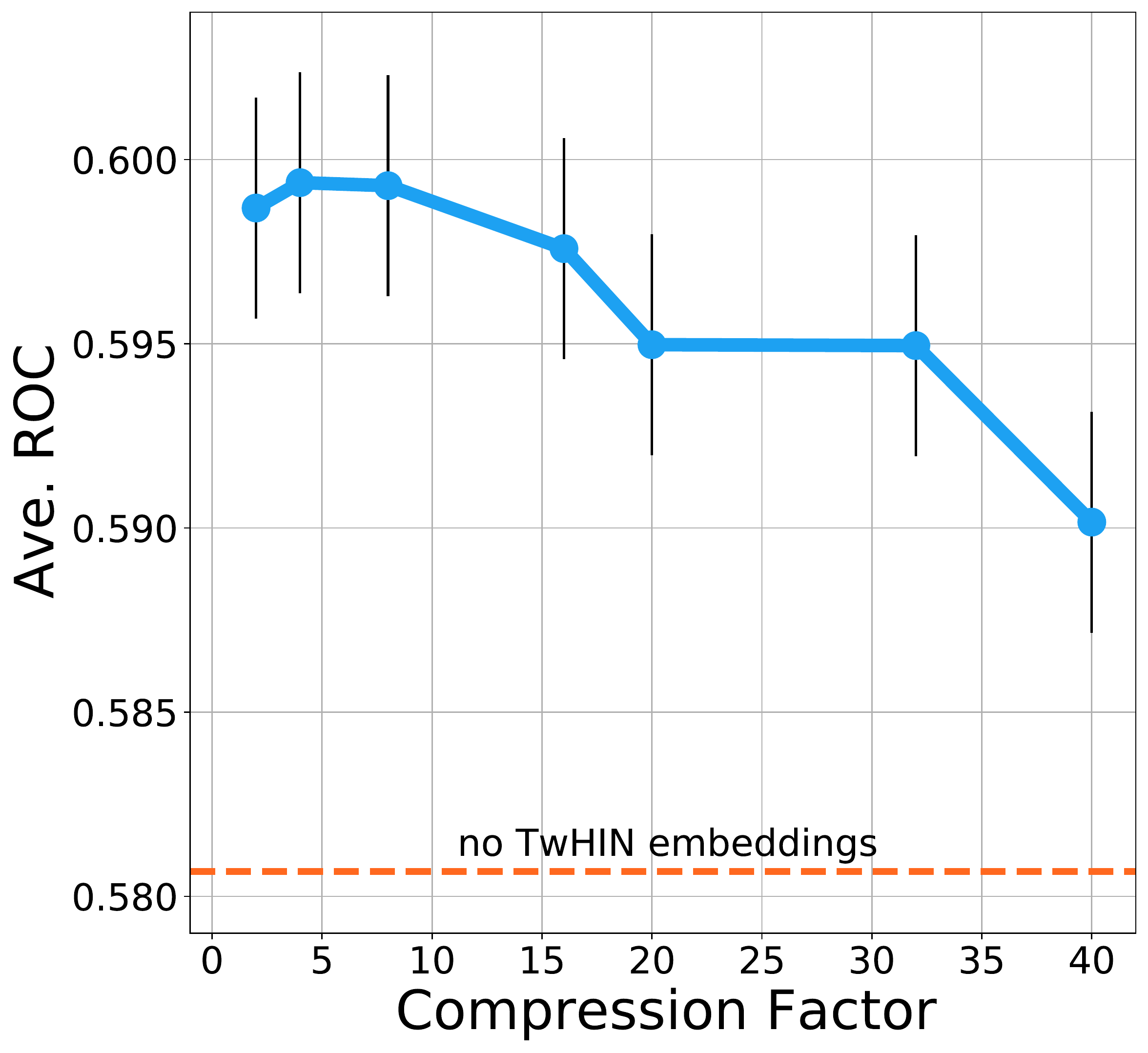}
         \caption{Ave. ROC for search ranking model lwith varying levels of compression for TwHIN.}
         \label{fig:ranking_compression}
     \end{subfigure}
     \hfill
     \begin{subfigure}[b]{0.315\textwidth}
         \centering
         \includegraphics[width=\textwidth]{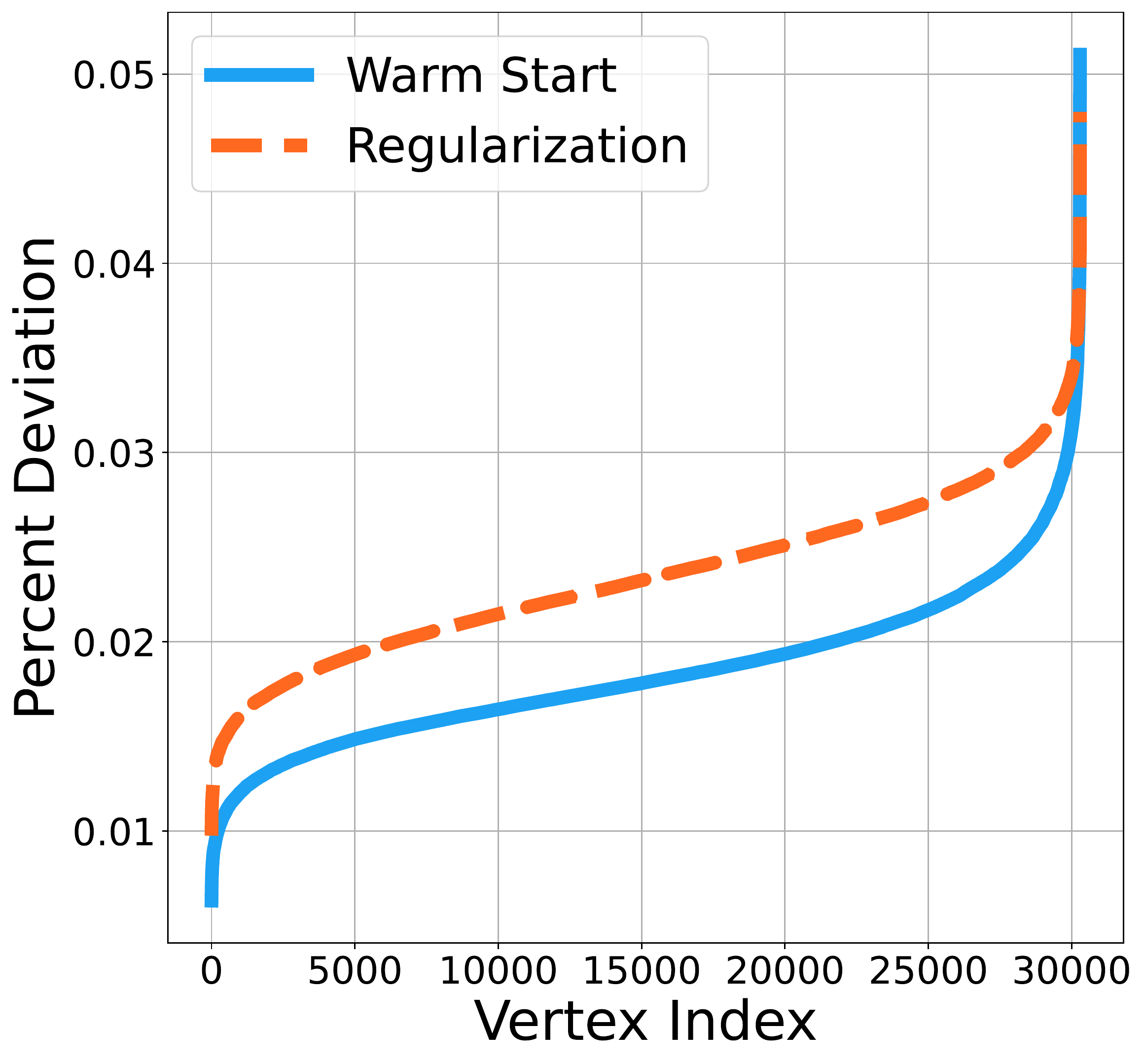}
         \caption{Deviation between consecutive TwHIN versions with parameter drift mitigation.}
         \label{fig:drift}
     \end{subfigure}
        \caption{Experiments exploring effects of compression on performance, and parameter drift mitigation strategies on drift.}
        \label{fig:three graphs}
\end{figure*}

\section{Practical Considerations}
\label{sec:design}
We discuss design decisions made in productionizing TwHIN with regards to (1) latency concerns and (2) mitigating technical debt by minimizing parameter drift.

\subsection{Compression for low latency}
 Performance in downstream ranking task often improves substantially as we increase the embedding dimension. Since, TwHIN embeddings are designed to be used in many latency-critical online recommender systems such as advertisement ranking and content ranking, we apply a simple and effective lossy data-based compression scheme via product quantization~\cite{jegou2010product}:
 \begin{enumerate}
     \item After generating the embeddings, we train a product quantization scheme (we use the FAISS package\footnote{\url{https://github.com/facebookresearch/faiss}}~\cite{JDH17}), and export the product quantization codebook (centers).
     \item Upon encountering a new downstream task, we make the codebook available to the training process, decoding the compressed embeddings using a codebook lookup.
     \item At inference time, we once again perform a codebook lookup.
 \end{enumerate}
 This scheme reduces the input size and network IO significantly, yields essentially identical downstream model performance, and introduces negligible latency. In Figure \ref{fig:knn_compression}, we consider the tradeoff between compression factor and the accuracy of decompressed embeddings in a $k$-nearest neighbors task. Even at high compression levels ($20\times$), we still maintain reasonable accuracy in the $k$-nearest neighbor task. Similarly, the effects of compression on the supervised search ranking task from Section~\ref{sec:exp_ranking}, show negligible effect on model performance even up to $30\times$ compression. These results motivate our approach to utilizing product quantization with model-side codebook decompression in our latency-critical recommendation tasks.

\subsection{Addressing parameter drift across versions}
Since the underlying information network contains user behaviors (e.g, follow or engagement actions) that evolve over time, TwHIN embeddings must be updated regularly to accurately represent entities. However, in doing so, we do not want the embeddings to drift too much from their current values, since we do want to simultaneously re-train all the downstream models.

Naively re-training the TwHIN embedding will lead to very large drifts caused by random initialization and stochasticity of optimization. In response, we have tried two natural approaches to achieving stability for embeddings: warm start and regularization. In the warm start approach, we simply initialize embeddings in the new version with the prior version's values. When we encounter new vertices that weren't previously seen, we either randomly assign them vectors, or initialize the vectors according to:
\begin{equation}
    \mathbf{\theta}_v = \frac{1}{\left| \mathcal{N}_v \right|} \sum_{v' \in \mathcal{N}_v} \mathbf{\theta}_{v'} + \theta_{\psi({v,v'})}
\end{equation}
where $\mathbf{\theta}_v$ is the embedding vector for node $v$, $\mathcal{N}_v$ is the graph neighborhood around $v$, and $\theta_{\psi{v,v'}}$, the relationship vector learned between vertices $v$ and $v'$.

Alternatively, adding regularization is a more principled way of addressing this issue, allowing us to directly penalize divergence from a past version. The simplest way is to apply L2 regularization to the previous embedding: $ \alpha \left| \left| \sum_v \left( \mathbf{\theta}_v - \mathbf{\theta}^{prev}_v \right) \right| \right|^2_2.$
Although this method is notionally simple, it presents the disadvantage of doubling the memory requirements, since we must also use $\theta^{prev}$.

We evaluate these two methods in terms of (1) parameter changes in L2 distance (2) the effect on downstream tasks. To assess parameter changes in L2 distance, we first generated a TwHIN embedding while optimizing for $30$ epochs. Afterwards, we re-trained for 5 epochs, separately applying warm-starting and L2 regularization. In Figure \ref{fig:drift}, warm-starting is better at minimizing deviations except in instances where the vertices have very high degree. Intuitively, this makes sense because the high-degree nodes are able to `overwhelm' the regularizer with their loss. Even still, a maximally 0.05\% deviation is more than sufficient to fulfill our stability requirements. 

\begin{table}[h]
\small
    \centering
    \begin{tabular}{c c c c}
    \toprule
    Metric & Control & Warm Start & Regularizer \\
    \midrule
    R@10 & 21.71 & \textbf{21.75} & 21.74 \\
    MRR & 0.0635 & 0.0635 & \textbf{0.0638} \\
    \bottomrule
    \end{tabular}
    \caption{Performance of various parameter-drift minimization strategies on Who to Follow task.}
    \label{tab:drift}
\end{table}
To evaluate the effect on downstream tasks, in Table~\ref{tab:drift}, we present a comparison on the Who to Follow task (Section~\ref{sec:cand_gen}). As we see in the above results, both warm start and regularization preserve stability in this downstream task. In practice, we have internally chosen to update TwHIN versions using the warm start strategy due to its space efficiency and simplicity.

%% file: 08conclusion.tex
\section{Conclusion}
\label{sec:conclusions}

In this work, we describe TwHIN, Twitter’s in-house joint embedding of multi-type, multi-relation networks with in-total over a billion nodes and hundreds of billions of edges. We posit that joint embeddings of heterogeneous nodes and relations is a superior paradigm over single relation embeddings to alleviate data sparsity issues and improve generalizability.  We demonstrate that simple, knowledge graph embedding techniques are suitable for large-scale heterogeneous social graph embeddings due to scalability and ease at incorporating heterogeneous relations. We deployed TwHIN at Twitter and evaluated the learned embeddings on a multitude of candidate generation and personalized ranking tasks. Offline and online A/B experiments demonstrate substantial improvements demonstrating the generality and utility of TwHIN embeddings. Finally, we detail many ``tricks-of-the-trade'' to effectively implement, deploy, and leverage large scale heterogeneous graph embeddings for many latency-critical recommendation and prediction tasks.